\documentclass[10pt,conference]{IEEEtran}
\IEEEoverridecommandlockouts
\usepackage{cite}
\usepackage{amsmath,amssymb,amsfonts}
\usepackage{algorithmic}
\usepackage{graphicx}
\usepackage{textcomp}
\usepackage{xcolor}
\usepackage{url}
\def\BibTeX{{\rm B\kern-.05em{\sc i\kern-.025em b}\kern-.08em
T\kern-.1667em\lower.7ex\hbox{E}\kern-.125emX}}

\usepackage{blindtext}
\usepackage{verbatim}
\usepackage{svg}
\usepackage{orcidlink}

\usepackage[braket, qm]{qcircuit}
\usepackage[]{algorithm2e}

\SetCommentSty{mycommfont}
\SetKwInput{KwInput}{Input}                
\SetKwInput{KwOutput}{Output} 
\makeatletter
\newcommand{\removelatexerror}{\let\@latex@error\@gobble}
\makeatother

\begin{document}

\title{Approximative lookup-tables and arbitrary function rotations for facilitating NISQ-implementations of the HHL and beyond
}
\author{\IEEEauthorblockN{1\textsuperscript{st} Petros Stougiannidis\textsuperscript{\textdagger}\thanks{\textsuperscript{\textdagger}These authors contributed equally.}$^{\orcidlink{0009-0000-3680-3655}}$}
\IEEEauthorblockA{\textit{LMU Munich}\\
petros.stougiannidis@gmail.com}
\and
\IEEEauthorblockN{2\textsuperscript{nd} Jonas Stein\textsuperscript{\textdagger}$^{\orcidlink{0000-0001-5727-9151}}$}
\IEEEauthorblockA{\textit{LMU Munich}\\
jonas.stein@ifi.lmu.de
}
\and
\IEEEauthorblockN{3\textsuperscript{rd} David Bucher$^{\orcidlink{0009-0002-0764-9606}}$}
\IEEEauthorblockA{\textit{Aqarios GmbH} \\
david.bucher@aqarios.com}
\and
\IEEEauthorblockN{4\textsuperscript{th} Sebastian Zielinski$^{\orcidlink{0009-0000-0894-8996}}$}
\IEEEauthorblockA{\textit{LMU Munich}\\
sebastain.zielinski@ifi.lmu.de}
\and
\IEEEauthorblockN{5\textsuperscript{th} Claudia Linnhoff-Popien$^{\orcidlink{0000-0001-6284-9286}}$}
\IEEEauthorblockA{\textit{LMU Munich}\\
linnhoff@ifi.lmu.de}
\and
\IEEEauthorblockN{6\textsuperscript{th} Sebastian Feld$^{\orcidlink{0000-0003-2782-1469}}$}
\IEEEauthorblockA{\textit{Delft University of Technology}\\
s.feld@tudelft.nl
}
}

\maketitle

\begin{abstract}
Many promising applications of quantum computing with a provable speedup center around the HHL algorithm. Due to restrictions on the hardware and its significant demand on qubits and gates in known implementations, its execution is prohibitive on near-term quantum computers. Aiming to facilitate such NISQ-implementations, we propose a novel circuit approximation technique that enhances the arithmetic subroutines in the HHL, which resemble a particularly resource-demanding component in small-scale settings. For this, we provide a description of the algorithmic implementation of space-efficient rotations of polynomial functions that do not demand explicit arithmetic calculations inside the quantum circuit. We show how these types of circuits can be reduced in depth by providing a simple and powerful approximation technique. Moreover, we provide an algorithm that converts lookup-tables for arbitrary function rotations into a structure that allows an application of the approximation technique. This allows implementing approximate rotation circuits for many polynomial and non-polynomial functions. Experimental results obtained for realistic early-application dimensions show significant improvements compared to the state-of-the-art, yielding small circuits while achieving good approximations.
\end{abstract}


\section{Introduction}
\label{sec:introduction}
The discovery of the HHL algorithm \cite{Harrow_2009} for solving linear systems of equations on quantum computers has opened up a plethora of applications for quantum computing while offering an exponential speedup over classical analogs under certain conditions \cite{Aaronson2015}. Unfortunately, its implementation as a quantum circuit is quite tedious due to necessary subroutines. Specifically, the eigenvalue inversion and the encoding of the resulting information into the amplitude of an ancilla qubit, which incorporates the evaluation of an arcsine function, pose a great challenge for NISQ-devices \cite{preskill_2018, yalovetzky2023nisqhhl}. While several approaches to this problem were proposed \cite{häner2018optimizing, cao2013poisson, yan2021module, Wang2020}, all of them require many ancilla qubits and high circuit depths. 

Moreover, there are approaches that have a worse-than-polynomial gate complexity, foremost lookup-tables that are implemented with uniformly controlled $R_y$ rotations \cite{Moettoenen_2004} and polynomial rotation circuits \cite{Qiskit} in combination with polynomial interpolation techniques. While these approaches are only viable for very small input sizes, both have the advantage that they are conceptually simple, easy to program and very space-efficient, i.e., they require no or only few ancilla qubits. Building on these favorable properties, this paper sets out to address the problem of exponential circuit depth by developing a procedure that can reduce the circuit depth of these circuits in a trade-off for accuracy. Concretely, the structure of polynomial rotation circuits allows evaluating the contribution of each individual gate to the overall result. We show that by omitting rotation gates with small contribution and high implementation cost, the circuits can be reduced in depth while the introduced error is held comparatively low. In addition, we show how lookup-tables, which in their canonical form have different structural properties, can be transformed such that their structure resembles that of the polynomial rotation circuits. Thereby, the approximation procedure is also made applicable to lookup-tables, which in return allows the compilation of approximate rotation circuits for non-polynomial functions.

Besides implementing rotations around the arcsine of a binary bit string, polynomial rotation circuits and lookup-tables can be used to rotate around any computable function. The ability to approximately implement arbitrary function rotations can be considered a fundamental quantum computing primitive, with potential applications in quantum algorithms similar to HHL, as well as in future quantum algorithms.

The structure of the paper is as follows: Sec.~\ref{sec:background} presents related work, followed by Sec.~\ref{sec:space_efficient_rotation}, which provides an algorithmic description of polynomial rotation circuits. Additionally, it provides a compilation algorithm that, given a polynomial $p$ as specification, compiles a corresponding quantum circuit that implements a rotation $R(p(x))$ for a binary number $x$, that is basis-encoded in a quantum register. Sec.~\ref{sec:polynomial_approximation} introduces a procedure that, given a preferred degree of approximation, reduces the circuit depth of polynomial rotation circuits by omitting rotation gates with a small contribution-to-cost ratio. In Sec.~\ref{sec:transfomring_lookup_tables}, an algorithm is presented that transforms the circuit structure of pre-compiled lookup-tables to resemble the structure of polynomial rotation circuits. The resulting approximate lookup-table circuits are evaluated in Sec.~\ref{sec:evaluation} by compiling circuits for different functions, input register sizes, and degrees of approximation, followed by numerical simulations to determine their accuracy, Toffoli gate counts, and required ancilla qubits.

\section{Background}
\label{sec:background}
\begin{figure*}[!t]
    \centering
\resizebox{0.8\textwidth}{!}{
\Qcircuit @C=1.0em @R=0.2em @!R { \\
	 	\nghost{{\mathrm{anc}} :  } & \lstick{{\mathrm{anc}} = \ket{0} :  } & \qw & \qw & \qw & \qw & \gate{R_y\,(\frac{1}{2^{n}})} & \qw & \push{\dots \quad} & \gate{R_y\,(\frac{1}{2})} & \qw & \qw & \qw  & \meter{1} & \rstick{1} \qw\\
	 	\nghost{{\mathrm{eig}}_{0} :  } & \lstick{{\mathrm{eig}}_{n} = \ket{0} :  } & \qw & \multigate{3}{\mathrm{QPE}(e^{iAt})} & \multigate{2}{\frac{1}{x}} & \multigate{2}{2\arcsin(Cx)} & \ctrl{-1} & \qw & \qw & \qw & \multigate{2}{2\arcsin(Cx)^\dagger} & \multigate{2}{\frac{1}{x}^\dagger} & \multigate{3}{\mathrm{QPE}(e^{iAt})^\dagger} & \qw & \rstick{\ket{0}} \qw\\
	 	\nghost{{\mathrm{eig}}_{1} :  } & \lstick{{\vdots}  } & \qw & \ghost{\mathrm{QPE}(e^{iAt})} & \ghost{\frac{1}{x}} & \ghost{2\arcsin(Cx)} & \qw & \qw & \push{\ddots \quad} & \qw &  \ghost{2\arcsin(Cx)^\dagger} & \ghost{\frac{1}{x}^\dagger} & \ghost{\mathrm{QPE}(e^{iAt})^\dagger}& \qw & \rstick{\ \vdots} \qw\\
	 	\nghost{{\mathrm{eig}}_{2} :  } & \lstick{{\mathrm{eig}}_{1} = \ket{0} :  } & \qw & \ghost{\mathrm{QPE}(e^{iAt})} & \ghost{\frac{1}{x}} & \ghost{2\arcsin(Cx)} & \qw & \qw & \qw &\ctrl{-3} & \ghost{2\arcsin(Cx)^\dagger} & \ghost{\frac{1}{x}^\dagger} & \ghost{\mathrm{QPE}(e^{iAt})^\dagger} & \qw & \rstick{\ket{0}} \qw \\
	 	\nghost{{\hat{b}}_{0} :  } & \lstick{{\hat{b}} :  } & {/} \qw & \ghost{\mathrm{QPE}(e^{iAt})} & \qw & \qw & \qw & \qw & \qw & \qw  & \qw & \qw & \ghost{\mathrm{QPE}(e^{iAt})^\dagger} & \qw & \rstick{C\hat{x}} \qw \\
\\ }
}
\caption{A high-level overview of the HHL algorithm, with each subroutine depicted as an independent module. The first module, $\mathrm{QPE}(e^{iAt})$, estimates the eigenvalues of the operator $e^{iAt}$, which encodes information about the matrix $A$ into the eigenvalue register $\ket{\mathrm{eig}}$. The second module, $\frac{1}{x}$, computes the reciprocal and can be implemented with an arithmetic circuit based on addition circuits~\cite{thapliyal2018quantum}. Finally, the third module, $2\arcsin(Cx)$, prepares the inverse eigenvalues to be encoded into $\ket{\mathrm{anc}}$ in a linearly-proportional fashion.} 
\label{fig:standard_hhl}
\end{figure*}
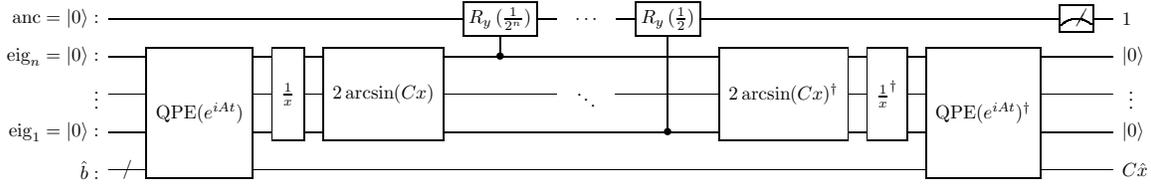

The HHL algorithm solves linear systems of equations $Ax = b$ in the sense that it prepares a quantum state $\hat{x}$ that is proportional to the classical solution vector $x$. For encoding the classical vectors $x$ and $b$ it uses amplitude encoding, e.g., $b = (b_1, \dotsc, b_j)^T$ is encoded in the amplitudes of a normalized quantum state
\begin{align}
    \hat{b} = \frac{(b_1, \dotsc, b_j)^T}{\sqrt{\sum_{k = 1}^{j} | b_k |^2}}.
\end{align} 
The HHL leverages the eigendecomposition of a Hermitian matrix $A$ and the Quantum Phase Estimation algorithm \cite{Cleve_1998} in order to efficiently compute a superposition of binary values \ket{\frac{1}{\lambda_j}} that are proportional to the inverse eigenvalues of $A$. Subsequently, this information needs to be encoded into the amplitudes of the quantum state $\hat{b}$, by introducing an ancilla qubit $\ket{\mathrm{anc}}$ and setting its amplitudes to $\ket{\mathrm{anc}} = \left( \sqrt{1 - {\frac{C^2}{{\lambda_j}^2}}} \; \frac{C}{\lambda_j} \right)^T$. The normalization constant $C$ needs to be chosen such that $\frac{C}{\lambda_j}$ do not exceed one in absolute value. Doing so manipulates the amplitudes of $\hat{b} \otimes \ket{\mathrm{anc}}$ such that the solution $\hat{x}$ is prepared into the subspace where $\ket{\mathrm{anc}}$ is in state $\ket{1}$:
\begin{align}
    \hat{b} \otimes  \begin{pmatrix} \sqrt{1 - {\frac{C^2}{{\lambda_j}^2}}} \\ \frac{C}{\lambda_j} \end{pmatrix} = (\dots ) \otimes \ket{0} + C \hat{x} \otimes \ket{1}.
\end{align}
The encoding of the rescaled inverse eigenvalues into the amplitudes of $\ket{\mathrm{anc}}$ can be implemented by deploying parameterized $R_y$ rotations that are controlled by the qubits of the register \ket{\frac{C}{\lambda_j}} \cite{cao2013poisson}. However, the relationship between the parameter $\theta$ and the induced amplitudes is non-linear:
 \begin{align}
	R_y(\theta) \ket{0} = \begin{pmatrix}
		\cos\left(\frac{\theta}{2}\right) \\[0.2cm]
		\sin\left(\frac{\theta}{2}\right)
	\end{pmatrix} = \cos\left( \frac{\theta}{2} \right) \ket{0} + \sin\left( \frac{\theta}{2} \right) \ket{1}.
\end{align}
Therefore, an additional computation needs to be performed, such that the amplitudes can be set linearly proportional to the inverse eigenvalues. For example, if the register that stores the inverse eigenvalues $\ket{\frac{C}{\lambda_j}}$ is transformed into $\ket{2\arcsin\left(\frac{C}{\lambda_j}\right)}$ beforehand, applying the $R_y$ rotations then yields
\begin{align}
    R_y\left(2\arcsin\left( \frac{C}{\lambda_j}\right)\right) \ket{0} = \sqrt{1 - {\frac{C^2}{{\lambda_j}^2}}} \ket{0} + \frac{C}{\lambda_j} \ket{1}.
\end{align}
Uncomputing the eigenvalue register and measuring a one in $\ket{\mathrm{anc}}$ results in the desired quantum state $\hat{x}$. A high-level overview of the quantum circuit can be seen in \figurename~\ref{fig:standard_hhl}.

However, computing one of the inverse trigonometric functions, arcsine or arccosine, to a high degree of accuracy is computationally expensive and, therefore, challenging to perform on NISQ devices. However, there is a simple and cost-efficient way to approximate this step using a first-order Taylor series approximation~\cite{cao2012quantum}: 
\begin{align}
    \arcsin(x) \approx x.
\end{align}
Such an approximation comes at no implementation cost as the circuit can directly rotate around the inverted eigenvalues, instead of deploying costly arithmetic circuits. This is possible due to the fact that the arcsine is almost linear close to zero. However, if the rescaled inverse eigenvalues are large in absolute value, a large error is introduced with this approximation. Although it is possible to further scale down the inverse eigenvalues to reduce the error, this approach negatively affects the post-selection step of the HHL algorithm and increases its average runtime \cite{cao2012quantum}. Since the arcsine function is monotonically increasing, decreasing its argument reduces the computed function value. Consequently, the induced amplitude in front of the $\ket{1}$ basis state will be smaller, making it less likely to measure the correct subspace.

In order to mitigate this problem, the arcsine can be approximated more accurately with polynomial interpolation techniques. Instead of the function of interest, one or more approximating polynomials are evaluated using arithmetic circuits for addition \cite{takahashi2009quantum,draper2000addition,Ruiz_Perez_2017, draper2004logarithmicdepth, cuccaro2004new}, multiplication \cite{muñozcoreas2017tcount,Ruiz_Perez_2017,häner2018optimizing} and the Horner scheme\footnote{The Horner scheme is an algorithm for evaluating polynomials of degree $d$ with $d$ additions and $d$ multiplications.}. The most prominent approach of this kind was proposed in~\cite{häner2018optimizing}, where a piecewise polynomial approximation circuit for high-accuracy function evaluations was implemented. However, the reversible nature of quantum circuits requires each intermediate result of the Horner evaluation to be stored in ancillary registers. This leads to high space-requirements when implemented as quantum circuits, with the number of ancilla qubits scaling linearly with the register size of the argument and the degree of the polynomial being evaluated. Iterative computations like the Horner scheme or ones that are based on Newton iterations can quickly reach ancilla requirements ranging from tens to hundreds of qubits \cite{häner2018optimizing, wiebe2014rus}.

In addition to explicit arithmetic implementations, there are also methods that allow for direct rotation around the desired value without the need for intermediate computations. One such method for rotating around the arcsine of an $n$-bit-value in a quantum register is to use precomputed lookup-tables that utilize $n$-fold controlled $R_y$ gates \cite{Moettoenen_2004}. These circuits are easy to program, space-efficient, and highly accurate. Their circuit depth, however, scales exponentially with $n$. Following this approach, a patent \cite{ibmpatent} uses lookup-tables in combination with an optimization technique called \textit{variable-sized binning}. Binning is used to reduce the exponential complexity with respect to the register size $n$ by treating growing batches of inputs as equivalent and approximating the computation. As a result, the computation scales linearly with $n$ but exponentially with the precision of the rotation. However, it is important to note that this approximation technique is only suitable for functions with a monotonically decreasing first derivative.

Another approach that utilizes precomputed rotation angles involves rotating around polynomial functions. The gate counts of these circuits are dominated by a sum of binomial coefficients $\sum_{k = 0}^{d} \binom{n}{d}$, where $n$ denotes the size of the quantum registers storing the argument and $d$ denotes the degree of the implemented polynomial. This sum of binomial coefficients, however, quickly becomes prohibitively large for even moderately large $n$ and $d$. Furthermore, this method introduces interpolation errors when approximating non-polynomial functions such as the arcsine using polynomials.

\section{Concept}
\label{sec:concept}
Aiming towards facilitating NISQ-implementations of quantum algorithms that incorporate arbitrary function rotations, we explore circuits for space-efficient polynomial rotations in Sec.~\ref{sec:space_efficient_rotation}. Afterwards, in Sec.~\ref{sec:polynomial_approximation}, we propose an efficient approximation of such circuits and introduce an approach to approximate lookup-tables in Sec.~\ref{sec:transfomring_lookup_tables}.

\subsection{Rotating around arbitrary polynomials space-efficiently}\label{sec:space_efficient_rotation}
The polynomial rotation circuits and a corresponding compilation algorithm can be derived by reformulating the mathematical expression for multiplying two $n$-bit integers $x = x_1 \dots x_n$ and $y = y_1 \dots y_n$ in terms of their binary representations 
\begin{align}
    x y = \sum_{i = 1}^{n}\sum_{j = 1}^{n} 2^{n-i} 2^{n-j} x_i  y_j.
\end{align}
Every summand is either zero or the product of the bit weights of the $i$-th bit of $x$ and the $j$-th bit of $y$ if both bits are set to one. This expression can be used to design a circuit for rotating around the product of $x$ and $y$:
\begin{align}
    R_y(xy)    &= R_y \left(\sum_{i = 1}^{n}\sum_{j = 1}^{n} 2^{2n-i-j} x_i  y_j \right).
\end{align}
Such a circuit can be implemented by preparing two input registers that store $x$ and $y$, and applying a doubly controlled rotation $C_{x_i}C_{y_j}R_y\left( 2^{2n-i-j} \right)$ for every $(i,j) \in \{ 1, \dotsc, n\}^2$ to the target qubit. Here, a unitary operation with a prefix $C_{x_i}$ denotes a controlled version of the operation in which the control is set on the qubit that stores the $i$-th bit of $x$. This idea can be generalized for rotating around the product of multiple factors, signed numbers in two's complement representation, and fixed-point fractional numbers. This ultimately allows for rotating around arbitrary monomials $ax^d$, where $a$ is a scalar coefficient, by preparing $d$ registers, each storing the argument $x$:
\begin{align}
    & R_y \left( a x^d \right) 
    =  R_y  \left( \sum_{i_1 = 1}^{n} \dots \sum_{i_d = 1}^{n}  a 2^{q - i_1} x_{i_1} \cdots 2^{q - i_d}x_{i_d} \right) \nonumber \\ 
    &= \prod_{\substack{(i_1, \dotsc, i_d) \\ \in \{1, \dotsc, n\}^d}} C_{x_{i_1}} \cdots C_{x_{i_d}} R_y  \left(a 2^{q - i_1} \cdots 2^{q - i_d} \right). 
\end{align}
Here, $q$ denotes the binary-point position of $x$%
\footnote{In general, it is also possible to choose bit weights such that the numbers that the argument register can represent are not equidistant, e.g., $x = x_1 \cdot -2^{-1} + x_2 \cdot 2^{-3} + x_3 \cdot 2^{-7} $.}. Note that, at this point, each control is set on exactly one qubit of a different input register, which all redundantly store $x$.

Concatenating several monomial rotations then allows for rotation around arbitrary polynomials. Evaluating a polynomial of degree $d$ on an $n$-bit argument results in $\mathcal{O}(n^{d})$ $d$-fold controlled rotation gates. Further, this requires $nd$ qubits for redundantly storing the argument and $d-1$ ancilla qubits for the $d$-fold controlled operations. We assume an implementation of multi-controlled single-qubit operations as described in \cite{nielsen2016}, which requires $2(k-1)$ Toffoli gates and $k-1$ ancilla qubits to implement a $k$-fold controlled gate. The ancilla qubits can be reused by subsequent rotation gates. As a result, the number of required ancilla qubits is determined by the rotation gate with the highest number of controls. 

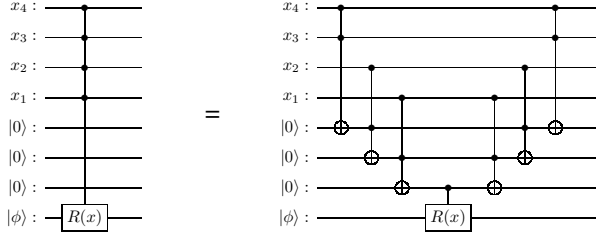
\begin{figure}[!t]
    \begin{minipage}[c]{0.35\columnwidth}
    \resizebox{!}{0.1cm}{
            \Qcircuit @C=1.0em @R=0.2em @!R { 
        	 	\nghost{{x\_1}_{4} :  } & \lstick{{x}_{4} :  } & \ctrl{1} & \qw & \qw\\
        	 	\nghost{{x\_1}_{3} :  } & \lstick{{x}_{3} :  } & \ctrl{1} & \qw & \qw\\
        	 	\nghost{{x\_1}_{2} :  } & \lstick{{x}_{2} :  } & \ctrl{1} & \qw & \qw\\
        	 	\nghost{{x\_1}_{1} :  } & \lstick{{x}_{1} :  } & \ctrl{4} & \qw & \qw\\
                \nghost{{x\_1}_{1} :  } & \lstick{\ket{0} :  } & \qw & \qw & \qw\\
                \nghost{{x\_1}_{1} :  } & \lstick{\ket{0} :  } & \qw & \qw & \qw\\
                \nghost{{x\_1}_{1} :  } & \lstick{\ket{0} :  } & \qw & \qw & \qw\\
        	 	\nghost{{anc} :  } & \lstick{{\ket{\phi}} :  } & \gate{R(x)} & \qw & \qw
            }   
        }
    \end{minipage}
    \begin{minipage}[c]{0.01\columnwidth}
        = 
    \end{minipage}
    \begin{minipage}[c]{0.35\columnwidth}
    \resizebox{!}{0.1cm}{
            \Qcircuit @C=1.0em @R=0.2em @!R { 
            	 	\nghost{{x\_1}_{4} :  } & \lstick{{x}_{4} :  } & \ctrl{1} & \qw & \qw & \qw & \qw & \qw & \ctrl{1} & \qw & \qw\\
            	 	\nghost{{x\_1}_{3} :  } & \lstick{{x}_{3} :  } & \ctrl{3} & \qw & \qw & \qw & \qw & \qw & \ctrl{3} & \qw & \qw\\
            	 	\nghost{{x\_1}_{2} :  } & \lstick{{x}_{2} :  } & \qw & \ctrl{2} & \qw & \qw & \qw & \ctrl{2} & \qw & \qw & \qw\\
            	 	\nghost{{x\_1}_{1} :  } & \lstick{{x}_{1} :  } & \qw & \qw & \ctrl{2} & \qw & \ctrl{2} & \qw & \qw & \qw & \qw\\
            	 	\nghost{{extra}_{0} :  } & \lstick{{\ket{0}} :  } & \targ & \ctrl{1} & \qw & \qw & \qw & \ctrl{1} & \targ & \qw & \qw\\
            	 	\nghost{{extra}_{1} :  } & \lstick{{\ket{0}} :  } & \qw & \targ & \ctrl{1} & \qw & \ctrl{1} & \targ & \qw & \qw & \qw\\
            	 	\nghost{{extra}_{2} :  } & \lstick{{\ket{0}} :  } & \qw & \qw & \targ & \ctrl{1} & \targ & \qw & \qw & \qw & \qw\\
            	 	\nghost{{anc} :  } & \lstick{{\ket{\phi}} :  } & \qw & \qw & \qw & \gate{R(x)} & \qw & \qw & \qw & \qw & \qw
            }}
            
    \end{minipage}
    \caption{Implementation of a four-fold controlled rotation gate. The right circuit shows how three ancilla qubits, initialized in state $\ket{0}$, can be utilized for implementing four-fold controlled operations.}
    \label{fig:k_control_decomposition}
\end{figure}

\figurename~\ref{fig:k_control_decomposition} depicts an implementation of a four-fold controlled rotation. \\
\IncMargin{1em}

\removelatexerror
\begin{algorithm}[H]

\DontPrintSemicolon
\LinesNumbered

\SetKwData{bits}{$X$}
\SetKwData{bitWeights}{$w$}
\SetKwData{polynomials}{$\mathrm{ps}$}
\SetKwData{polynomial}{$p$}
\SetKwData{coefficients}{$a$}
\SetKwData{degree}{$d$}
\SetKwData{gate}{$\left( c_1, \dotsc, c_d \right)$}
\SetKwData{controlqubits}{$\mathrm{cq}$}
\SetKwData{rotationvalue}{$\theta$}
\SetKwArray{circuit}{$\mathrm{circuit}$}
\SetKwFunction{accumulate}{accumulate}
\SetKwFunction{dict}{HashMap}
\SetKwFunction{set}{HashSet}

\SetKwInOut{Input}{input}
\SetKwInOut{Output}{output}

\Input{Register size $n$, a list of coefficients $\coefficients = [a_0, \dotsc, a_d]$ that specifies a polynomial $p = \sum_{i = 0}^{d} a_i x^i$ and a list \bitWeights that stores the weights of the argument bits.}
\Output{A map \circuit that assigns a set of control qubits in the input register to a rotation angle according to the input polynomial.}
\BlankLine
\bits $\gets \{0, \dotsc, n-1\}$\;
\circuit $\gets \dict{}$\; 
    \ForAll{$\degree \in \{ 0, \dotsc, \deg(\polynomial)\}$}{
        \ForAll{$\gate \in \bits^{\degree}$}{
            \rotationvalue $\gets \coefficients_{\degree}$\;
            \For{$i \gets 1$ \KwTo \degree}{
            \rotationvalue $\gets \rotationvalue \cdot \bitWeights_{c_i}$\;
            }
            \controlqubits $\gets \set(c_1, \dotsc, c_d )$\;
            \eIf{\controlqubits $\in$ \circuit}{
                $\circuit[\controlqubits] \gets \circuit[\controlqubits] + \rotationvalue$
            }
            {
               $\circuit[\controlqubits] \gets \rotationvalue$ 
            }
        }
    }
\BlankLine
\caption{A naive $\mathcal{O}(n^d)$ algorithm for compiling polynomial rotation circuits.}\label{alg:compilation}
\end{algorithm}
\DecMargin{1em} 
However, as this circuit is derived from generic multiplication rather than exponentiation, it can be optimized by using the assumption that every argument of the multiplication is the same factor $x$. Thereby, almost all rotation gates of the circuit can be replaced by  gates with fewer controls. Most importantly, this optimization circumvents the need to prepare the $n$-qubit registers $d-1$ times, which store $x$ redundantly. Hence, the ancilla requirement is reduced from $n(d-1) + (d-1)$ to $d-1$. Additionally, sets of rotation gates that share the same control qubits can be collapsed into a single rotation. Since there are only $2^n$ possible constellations of control qubits on a register of size $n$, the number of rotation gates is limited to $2^n$, and the ancilla requirements are additionally bound by $n-1$. This optimization task can be performed by a classical algorithm that takes a specification of a polynomial as input and compiles an optimized polynomial rotation circuit (see Algorithm \ref{alg:compilation}).

The naive implementation of the algorithm scales in $\mathcal{O}(n^d)$ as it exhaustively inspects every rotation, determines its rotation value (lines 5-8), and the set of bits of $x$ it is ultimately controlled by (line 9). The rotation value is then accumulated into a rotation gate of the new circuit that is controlled by exactly the determined set of bits. The general structure of the resulting polynomial rotation circuits is shown in \figurename~\ref{fig:transformed_circuit}.

Furthermore, inspecting the compilation procedure reveals that if $n$ is chosen larger than $d$, a certain subset of the $2^n$ possible compiled rotation gates are guaranteed to have a rotation angle of zero and can therefore be omitted. Precisely, any gate with a number of control qubits greater than $d$ will never accumulate any rotation angle because a $d$-fold controlled rotation can only be controlled by at most $d$ of the $n$ bits of $x$. Line 9 in Algorithm \ref{alg:compilation} can, therefore, never produce any set with a cardinality larger than $d$. By properly implementing the compilation, i.e., not initializing such gates, the resulting circuit consists of at most $\min\left(2^n, \sum_{i = 0}^{d} \binom{n}{d}\right)$ rotation gates and $\min(n,d) - 1$ ancilla qubits next to argument register. Tab.~\ref{tab:upper_bounds} gives an overview of these circuit properties.

\begin{table}
\begin{center}
\caption{The number of rotation gates, ancilla qubits, and Toffoli gates for different $n$ and $d$. The computation of the Toffoli and ancilla count assumes an implementation of the multiple-controlled rotation gates as described in~\cite{nielsen2016}.}
\label{tab:upper_bounds}
\renewcommand{\arraystretch}{1.3}
\begin{tabular}{ c | c | c }
    & $n \leq d$  & $n > d$ \\ \hline
    Rotation gates & $2^n$ & $\sum_{k=0}^{d} \binom{n}{d}$ \\ \
    Ancilla qubits & $n-1$ & $d - 1$ \\ \
    Tofolli gates & $\sum_{k = 1}^{n} \binom{n}{k} \cdot 2(k-1)$ & $\sum_{k = 1}^{d} \binom{n}{k} \cdot 2(k-1)$ 
\end{tabular}
\renewcommand{\arraystretch}{1}
\end{center}
\end{table}

To our knowledge, there is no published literature on this approach. However, a more sophisticated algorithm for compiling polynomial rotation circuits is implemented in the \texttt{PolynomialPauliRotations} module in IBM's Qiskit~\cite{Qiskit}. This algorithm first prepares the $\min\left(2^n, \sum_{i = 0}^{d} \binom{n}{d}\right)$ final rotation gates and subsequently uses multinomial coefficients in order to compute their rotation values more efficiently. 

\subsection{Introducing approximate polynomial rotation circuits} \label{sec:polynomial_approximation}
Despite the fact that the presented polynomial rotation circuits are highly space-efficient, the circuit depth becomes prohibitive for even moderate input sizes. Other state-of-the-art approaches typically reach smaller circuit depths while keeping approximation errors in acceptable regions. For example, binning \cite{ibmpatent} or truncated multiplication \cite{häner2018optimizing} introduce a certain error to the computation in a trade-off for smaller circuits. In contrast, the polynomial rotation circuit, as described up to this point, computes the polynomial rotation exactly, i.e., there is no approximation and thus no error. In order to further reduce the circuit depth, additional gates could be omitted. However, as all remaining gates have non-zero rotation angles, the final polynomial rotation will inevitably become approximate. If further gates were to be omitted, there should be a worthwhile trade-off between circuit size and introduced error. Fortunately, compiling and inspecting these circuits very often reveals rotation gates with minuscule rotation angles, especially when the coefficients and the degree of the polynomial are not too large. 


In the following, we denote a rotation gate with a rotation value of $\theta$ and controlled by the set of qubits $s$ by $(\theta, s)$. In general, rotation gates with small $\theta$ contribute less to the final rotation, i.e., the absolute error introduced by omitting these gates is small compared to gates with large angles. Moreover, rotation gates with many control qubits are also worthy candidates for omission as they are more expensive to implement than gates with fewer control qubits. Naturally, with every omitted gate the potential error further increases. In order to upper-bound the introduced error for a certain set of omitted gates $O$, consider a circuit $R_p$ that implements a rotation around a polynomial $p$ and is applied on a register storing a certain value $x$. Each rotation gate in $R_p$ is controlled by a certain set of qubits of said register, and depending on which qubits are in state $\ket{1}$, some gates will contribute to the final rotation, while others will not. If $O$ is omitted from $R_p$, where $R_p = \tilde{R}_p \cup O$ and $\tilde{R}_p \cap O = \emptyset$, the introduced error in the resulting approximate circuit $\tilde{R}_p$ is given by:
\begin{align}
    &\left| \sum_{\substack{(s, \theta) \in R_p}} \theta \prod_{i \in s} x_i \ -  \sum_{\substack{(s, \theta) \in \tilde{R}_p}} \theta \prod_{i \in s} x_i  \right| = \nonumber \\
    %
    %
    &\left|\sum_{(s, \theta) \in O} \theta \prod_{i \in s} x_i \right| \leq \sum_{(s, \theta) \in O} |\theta| \prod_{i \in s} x_i \leq \sum_{(s, \theta) \in O} |\theta|.
\end{align}
The worst-case scenario is then easily identified to be when $x$ is a bit string consisting purely of ones. In this case, each gate contributes to the final result. The largest possible deviation from the correct result can therefore be upper bounded by the sum of the absolute values of the rotation angles of all gates in $O$.

A straightforward approach to further shrink the circuit depth is to keep omitting gates until their absolute rotation values add up to a threshold, namely the maximum error one is willing to accept. In order to minimize the gate count, rotation gates with small rotation angles should be omitted first. However, in order to minimize the Toffoli count, the decision on which gates to omit first should not only be based on the absolute value of their rotation angles, but also on the number of required Toffoli gates. Hence, the compiled rotation gates can be sorted by a ratio of rotation value and Toffoli count before gates are omitted until the threshold is reached. An alternative approach is to omit rotation gates from the sorted list until a certain circuit depth is reached, and then analyze the introduced error.

The approximation procedure not only reduces the circuit depth but also has the potential to decrease the number of required ancilla qubits. For instance, omitting all $k$-fold controlled rotation gates, where $k$ is the largest number of control qubits across all rotation gates of a circuit, leads to a reduction in the ancilla count from $k-1$ to $k-2$. In some cases, the dimensions of a polynomial rotation circuit can be significantly reduced while still maintaining satisfactory accuracies. For instance, an exact polynomial rotation circuit implementing $R(x^7)$, where $-0.5 \leq x < 0.5$ is stored in a $14$-qubit register, requires 94874 Toffoli gates, whereas an approximate circuit that implements the rotation with a maximum error of $3.01 \times 10^{-5}$ across all possible inputs $x$ requires only 4348 Toffoli gates. A circuit with a maximum error of $2.93 \times 10^{-4}$ needs only 1298 Toffoli gates. Depending on the implemented polynomial functions, the speed-ups can be smaller or larger. Ultimately, this makes this approach very promising for expanding the use cases of polynomial rotation circuit beyond toy problems.

\subsection{Transforming the structure of lookup-tables into the structure of polynomial rotation circuits} \label{sec:transfomring_lookup_tables}
\begin{figure}[!t]
\hspace{0cm}
    \resizebox{0.8\columnwidth}{!}{
        \Qcircuit @C=1.0em @R=0.2em @!R { 
    	 	\nghost{{x}_{3} :  } & \lstick{{x}_{3} :  } & \ctrlo{1} & \ctrl{1} & \qw & \push{\phantom{\dots} \quad} & \ctrl{1} & \qw\\
    	 	\nghost{{x}_{2} :  } & \lstick{{x}_{2} :  } & \ctrlo{1} & \ctrlo{1} & \qw & \push{\phantom{\dots} \quad} & \ctrl{1}  & \qw  \\
    	 	\nghost{{x}_{1} :  } & \lstick{{x}_{1} :  } & \ctrlo{1} & \ctrlo{1} & \qw & \push{\dots \quad} & \ctrl{1} & \qw \\
    	 	\nghost{{anc} :  } & \lstick{{\ket{\phi}} :  } & \gate{\mathrm{R_y(f(000_b))}} & \gate{\mathrm{R_y(f(001_b))}} & \qw & \push{\phantom{\dots} \quad}  & \gate{\mathrm{R_y(f(111_b))}}  & \rstick{R_y \left( f(x) \right) \ket{\phi}} \qw \\
        } 
    } 
\caption{The general structure of a lookup-table implementing a rotation around a function $f$. Here, exactly one rotation gate makes the entire contribution to the final result for a certain input $x$. White bullets indicate a control on a qubit in state $\ket{0}$, black bullets control on a qubit in state $\ket{1}$.}
\label{fig:LUT}
\end{figure}
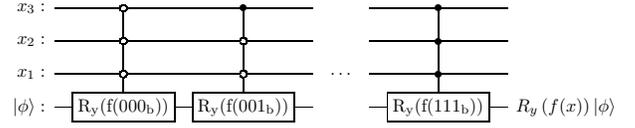

Lookup-tables and polynomial rotation circuits differ in their structure, particularly in how their gates contribute to the final result. In a lookup-table, only one of its $2^n$ gates performs a non-identity operation in each evaluation (assuming that the qubits of the argument register are only in states $\ket{0}$ or $\ket{1}$), while in a polynomial rotation circuit, multiple gates are involved. Since the single gate in a lookup-table is essential to the computation, omitting any subset of gates is not feasible. Thus, the approximation procedure presented in Sec.~\ref{sec:polynomial_approximation} is not applicable to lookup-tables due to their distinct properties.
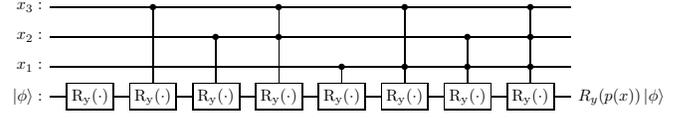
\begin{figure}[!t]
\hspace{-0.5cm}
    \resizebox{0.9\columnwidth}{!}{
        \Qcircuit @C=1.0em @R=0.2em @!R { 
        	 	\nghost{{x\_1}_{0} :  } & \lstick{{x}_{3} :  } & \qw & \ctrl{3} & \qw & \ctrl{1} & \qw & \ctrl{2} & \qw & \ctrl{1} & \qw\\
        	 	\nghost{{x\_1}_{1} :  } & \lstick{{x}_{2} :  } & \qw & \qw & \ctrl{2} & \ctrl{2} & \qw & \qw & \ctrl{1} & \ctrl{1} & \qw\\
        	 	\nghost{{x\_1}_{2} :  } & \lstick{{x}_{1} :  } & \qw & \qw & \qw & \qw & \ctrl{1} & \ctrl{1} & \ctrl{1} & \ctrl{1} & \qw\\
        	 	\nghost{{anc} :  } & \lstick{{\ket{\phi}} :  } & \gate{\mathrm{R_y(\cdot)}} & \gate{\mathrm{R_y(\cdot)}} & \gate{\mathrm{R_y(\cdot)}} & \gate{\mathrm{R_y(\cdot)}} & \gate{\mathrm{R_y(\cdot)}} & \gate{\mathrm{R_y(\cdot)}} & \gate{\mathrm{R_y(\cdot)}} & \gate{\mathrm{R_y(\cdot)}} & \rstick{R_y(p(x)) \ket{\phi}} \qw
        }
    }
    \caption{The general structure of polynomial rotation circuits, implementing a rotation around a polynomial $p$. Here, a rotation gate makes a contribution to the final result if its set of control qubits is a subset of the set of the input bits that are in state $\ket{1}$.}
    \label{fig:transformed_circuit}
\end{figure}
However, extending the approximation procedure to include lookup-tables would be highly desirable, as lookup-tables can compute any computable function. In the context of the HHL algorithm, this would allow to implement approximate arcsine rotations, which would eliminate the need for polynomial interpolation and the associated interpolation error. For this reason, we developed an algorithm that transforms a compiled lookup-table, as depicted in Fig.~\ref{fig:LUT}, into circuits with the same structure as polynomial rotation circuits (see Fig.~\ref{fig:transformed_circuit}), allowing for potential reduction in circuit depth. First, our algorithm modifies the control mechanism of rotation gates of the lookup-table by removing the controls on the $\ket{0}$ states. By removing the restrictive $\ket{0}$-controls, more rotation gates perform a non-identity operations during an evaluation. However, in order to restore the correctness of the lookup-table the rotation values of the gates need to be adjusted in order to compensate for the changes. Therefore, the algorithm subsequently inspects each rotation gate $(s_1,\theta_1)$ and subtracts its rotation value $\theta_1$ from the rotation values $\theta_2$ of each rotation gate $(s_2, \theta_2)$ whose set of control qubits $s_2$ are a superset of $s_1$ (see Algorithm \ref{alg:LUT_transformation}). This ensures that the rotation values of gates that perform a non-identity operation for a certain input add up to the correct final rotation angle.


\removelatexerror
\begin{algorithm}[H]

\DontPrintSemicolon
\LinesNumbered

\SetKw{KwWhere}{where}

\SetKwData{circuit}{$\mathrm{circuit}$}
\SetKwData{bits}{$X$}
\SetKwData{bit}{$i$}
\SetKwData{function}{$f$}
\SetKwData{bitWeights}{$w$}
\SetKwData{ss}{$\mathrm{s_1}$}
\SetKwData{ls}{$\mathrm{s_2}$}

\SetKwData{value}{$x$}
\SetKwData{input}{$s$}
\SetKwFunction{dict}{HashMap}

\SetKwInOut{Input}{input}
\SetKwInOut{Output}{output}

\Input{Register size $n$, a function \function and a list \bitWeights that stores the weights of the argument bits.}
\Output{A map \circuit that assigns a set of control qubits in the input register to a rotation angle according to the input function.}
\BlankLine

\bits $\gets \{ 0, \dotsc, n-1\}$\;
\circuit $\gets \dict{}$\;
\ForAll{$\input \in \mathcal{P}(\bits)$}{
    \value $\gets 0$\;
    \ForAll{$\bit \in \input$}{
        \value $\gets \value + \bitWeights_{\bit}$\;
    }
    $\circuit[\input] =  \function(\value)$\;
}

\For{$i \gets 0$ \KwTo $n-1$ }{
   \ForAll{\ss $\in \mathcal{P}(\bits)$ \KwWhere $|\ss| = i$}{
        \ForAll{\ls $\in \mathcal{P}(\bits)$ \KwWhere $|\ls| > i$}{
            \If{$\ss \subset \ls$}{
                $\circuit[\ls] \gets \circuit[\ls] - \circuit[\ss]$\;
            }
        }
    }
}

\BlankLine
\caption{An $\mathcal{O}(n2^{2n})$ algorithm for compiling and transforming the structure of a lookup-table into the structure of polynomial rotation circuits.}\label{alg:LUT_transformation}
\end{algorithm}
Overall, in the modified lookup-tables, the contribution of a single rotation gate is distributed among many others. This transformation has the advantage that lookup-tables can now be made subject to the proposed approximation procedure, as it is now possible to assess each rotation gate regarding its contribution to the final result and its implementation cost. Consequently, this allows for compiling approximate rotation circuits for any computable function, not only polynomials. The efficiency of the approximation, however, heavily depends on the function to be implemented.

In addition, this algorithm is capable of compiling the exact same circuits as the polynomial compilation algorithms, up to rounding errors and zero-angle rotation gates that can simply be filtered out. If the transformed lookup-tables implement polynomial functions, their circuit dimensions are also bound by the formulas in Tab.~\ref{tab:upper_bounds}. This behavior is also observed in our experiments (see Sec.~\ref{sec:evaluation}). In that sense, the algorithm offers a more generalized approach to the compilation of both lookup-tables and polynomial rotation circuits.

Due to its exponential scaling with register size but independence of polynomial degree, this algorithm is best suited for compiling circuits for high-degree polynomials and non-polynomial functions when the register sizes are moderate\footnote{For $n \leq 17$, the compilation times range from microseconds to a few minutes. However, for $n = 18$, the compilation time became excessively long, taking around one hour on a conventional computer with an Intel Core i7-8550U CPU and 16 GB of RAM.}. For this reason, the Qiskit algorithm remains relevant for cases where low-degree polynomial circuits need to be compiled for large register sizes.

\section{Evaluation}
\label{sec:evaluation}
In order to evaluate the space and time efficiency, as well as the accuracy, of the proposed approximative lookup-table approach, we now conduct numerical simulations. For that, lookup-tables are compiled for various functions, argument register sizes, and degrees of approximation to assess their performance.

\subsection{Strategy for compiling and approximating lookup-tables}\label{sec:strategy}
First, lookup-tables are compiled and their structure is transformed with a Python implementation\footnote{Implementation source code\\\url{https://github.com/petros-stougiannidis/quantum-rotation-compiler}.} of Algorithm \ref{alg:LUT_transformation}. The bit weights of the argument registers of size $n$ are chosen such that they represent $2^n$ values in a certain interval. For instance, choosing $w = [-0.5, 0.25, 0.125, \dotsc, 2^{-n} ]$ lets the argument register represent $2^n$ values in the interval $[-0.5, 0.5[$ in two's complement representation. Then, the rotation gates within each circuit, which cost at least one Toffoli gate, are sorted according to a contribution-to-cost ratio. For a rotation gate $(s,\theta)$, the ratio gets computed with:
\begin{align}
   \frac{ \left| \theta \right|}{2(|s|-1)},
\end{align}
where the set $s \in \mathcal{P}(\{0, \dotsc, n-1\})$ denotes the indices of its control qubits, and $\theta$ is the rotation value represented as a double-precision floating point number. The denominator computes the Toffoli count of a rotation gate depending on the number of its controls. This heuristic was chosen in order to encapsulate the contribution to the final result as well as the implementation cost: Thereby, the approximation procedure is guided towards a reduction of the Toffoli count. Subsequently, the rotation gates with the lowest contribution-to-cost ratio are removed until the Toffoli count of the remaining gates falls below a set threshold.

\subsection{Assessing the accuracy of a lookup-table}\label{sec:measure_of_accuracy}
To evaluate the accuracy of each circuit, a numerical simulation is performed on all possible quasi-classical states of the argument register, i.e., where each qubit is limited to the states $\ket{0}$ or $\ket{1}$. There are $2^n$ such inputs for a register size of $n$, which are specified by a set containing the indices of all qubits in state $\ket{1}$. For each input, the rotation gates in the approximate circuit are iterated. Whenever the set of control qubits $s$ of a rotation gate resembles a subset of the input qubit indices, the rotation value $\theta$ is added to an accumulator variable. The value of the accumulator is then treated as the simulated output of the circuit for the given input. The accuracy of the circuit is measured by computing the absolute error as the distance between the simulated output $\tilde{f}(x)$ and the optimal output $f(x)$, and then selecting the largest absolute error across all inputs:
\begin{align}
    \max_{x \in X} \left| f(x) - \tilde{f}(x)  \right|,
    \label{accuracy}
\end{align}
where $X$ denotes all $2^n$ input values. The average errors are computed as
\begin{align}
    \frac{1}{|X|}\sum_{x \in X} \left| f(x) - \tilde{f}(x) \right|.
\end{align}

\subsection{Evaluating gate efficiency and approximation accuracy}\label{sec:results}
\begin{figure}[!t]
    \centering
    \includegraphics[width=\columnwidth]{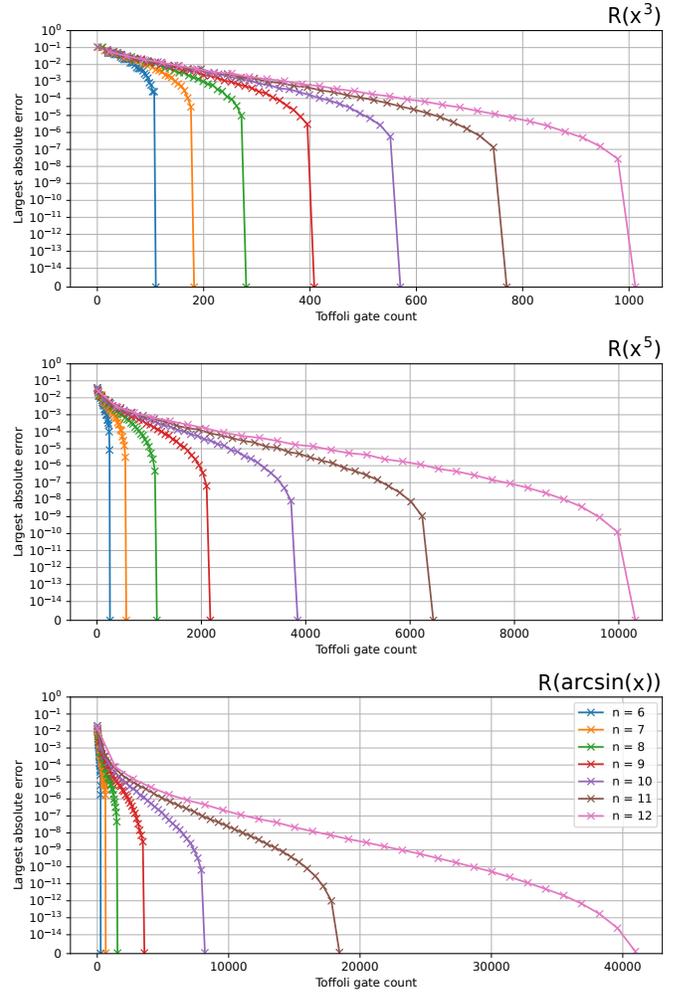}
    \caption{The accuracies of approximate lookup-tables compiled for the functions $x^3$ (top), $x^5$ (middle), $\arcsin(x)$ (bottom) for different circuit depths (x-axis) and argument registers sizes $n$ (colored). Each circuit is evaluated on $2^n$ inputs in the interval $[-0.5, 0.5[$.}
    \label{fig:approximation_efficiency}
\end{figure}

The three plots in \figurename~\ref{fig:approximation_efficiency} depict the performance of lookup-tables implementing $R(x^3)$, $R(x^5)$ and $R(\arcsin(x))$, respectively. The circuits were evaluated for values in the interval $[-0.5, 0.5[$. Each data point represents the accuracy of a different lookup-table, with the y-axis showing the largest error across all inputs (see Eq.~\ref{accuracy}) and the x-axis showing the Toffoli count of the circuit. The better a lookup-table performs on implementing the desired approximate rotation, the lower its error on the y-axis and the lower its Toffoli count on the x-axis. Lookup-tables that were compiled and simulated for the same register size are color-coded identically. The colored graphs for a fixed register size show the error that is introduced when reducing the circuit depth with the approximation procedure from Sec.~\ref{sec:polynomial_approximation}. Flatter sections of these graphs correspond to lower introduced error when omitting further rotation gates.

Similarly, steep slopes in a graph indicate that a circuit greatly profits in terms of accuracy with more gates invested. For example, the plot of the $R(\arcsin(x))$ rotations displays steep slopes in the region of the x-axis between 0 and approximately 5000 Toffoli gates. Investing a number of Toffoli gates corresponding to these regions of the x-axis yields the highest increase in accuracy. Data points in these regions therefore highlight sweet spots for compiling shallow circuits with comparatively high accuracy. Tab.~\ref{tab:nisq_circuit_sizes_and_accuracies} shows a few selected example circuits for the $R(\arcsin(x))$ rotations that achieved moderately high accuracy (in the order of $10^{-3}$ to $10^{-6}$), with a reasonable circuit depth (between 98 and 1298 Toffoli gates) and remarkably small circuit width (2-6 ancilla qubits). For comparison, the accuracy of a first-order Taylor series approximation in the interval $[-0.5, 0.5[$ is $2.36\times 10^{-2}$. Therefore, investing a few hundred Toffoli gates and a small number of ancilla qubits increases the accuracy by several orders. Moreover, our approach compares favorably to \cite{häner2018optimizing} in terms of ancilla requirements for implementing arbitrary function rotations. The authors did not evaluate the accuracy of their approach for register sizes smaller than 30. However, implementing piecewise-linear interpolation with their approach requires at least 17, 21 and 25 ancilla qubits for $n = 8$, $n = 10$ and $n = 12$, respectively. Implementing piecewise-cubic interpolation already results in ancilla requirements of 33, 41, and 49 for these register sizes. Hence, in these smaller-scale settings and specifically for implementing function rotations, our approach shows a noticeable improvement.

\begin{figure}[!t]
    \centering
    \includegraphics[width=\columnwidth]{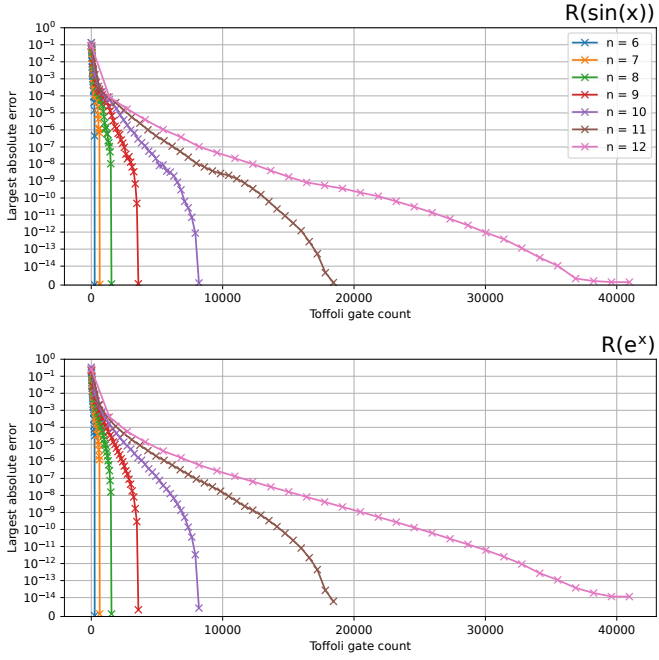}
    \caption{The accuracies of approximate lookup-tables compiled for the functions $\sin(x)$ (top), $e^x$ (bottom) for different circuit depths (x-axis) and argument registers sizes $n$ (colored). Each circuit is evaluated on $2^n$ inputs in the interval $[-1, 1[$.}
    \label{fig:approximation_efficiency_nonpols}
\end{figure}

Fig.~\ref{fig:approximation_efficiency} also highlights an interesting phenomenon: the sudden decrease in slopes at certain circuit depths. The intersections of the graphs with the x-axis signify the point beyond which adding more depth to the circuit no longer improves accuracy. Ideally, it's best to reach these points with small circuit depths, as that implies highly accurate rotations with low implementation costs. However, the circuit depth at which a graph intersects with the x-axis can vary substantially depending on the implemented function\footnote{Note that the x-axis scales differently across the plots for $R(x^3), R(x^5)$ and $R(\arcsin(x))$.}. For instance, the simulations demonstrate that polynomial functions of lower degree hit this point earlier than those of higher degree. Specifically, these circuit depths coincide with the Toffoli counts of the polynomial rotation circuits outlined in Tab.~\ref{tab:upper_bounds} in Sec.~\ref{sec:polynomial_approximation}. In contrast, polynomial functions of degree $d \geq n$ and non-polynomial functions such as $\arcsin(x)$, $e^{x}$, and $\sin(x)$ (see Fig.~\ref{fig:approximation_efficiency_nonpols}) do not intersect the x-axis unless all $2^n$ rotation gates are employed. Nevertheless, it's important to note that even if a circuit reaches the highest possible accuracy, there may still be noticeable rounding errors introduced by floating point numbers and a large number of additions during circuit compilation and simulation. This effect is especially noticeable in the $R(e^x)$ rotations of Fig.~\ref{fig:approximation_efficiency_nonpols}, where machine precision ($\approx 10^{-16}$) cannot be reached, even when all $2^n$ rotation gates are used.

\begin{table}
\centering
\caption{Toffoli gate count, ancilla requirements and errors for approximate circuits implementing $R(\arcsin(x))$, where the argument $x$ is stored in an $n$-qubit register. The circuits were reduced in depth up to a maximal Toffoli count of 100, 500, 900 and 1300, respectively.} 
\label{tab:nisq_circuit_sizes_and_accuracies}
\begin{tabular}{c|c|c|c|c}
& Toffoli & Ancilla & Average error & Largest error \\
\hline
$n=8$ & 100 & 2 & 4.54e-04 & 3.33e-03 \\ 
& 494 & 4 & 1.46e-05 & 1.62e-04 \\ 
& 894 & 5 & 5.67e-07 & 1.41e-05 \\ 
& 1292 & 6 & 3.61e-08 & 1.19e-06 \\ 
\hline
$n=10$ & 98 & 2 & 4.58e-04 & 3.44e-03 \\ 
& 498 & 4 & 3.55e-05 & 3.47e-04 \\ 
& 896 & 4 & 8.89e-06 & 1.13e-04 \\ 
& 1298 & 4 & 2.84e-06 & 4.21e-05 \\ 
\hline
$n=12$ & 98 & 2 & 4.66e-04 & 3.56e-03 \\ 
& 496 & 4 & 5.87e-05 & 5.04e-04 \\ 
& 896 & 4 & 1.67e-05 & 1.79e-04 \\ 
& 1294 & 4 & 6.83e-06 & 8.67e-05 \\
\end{tabular}
\end{table}

\begin{figure}[!t]
    \centering
    \includegraphics[width=\columnwidth]{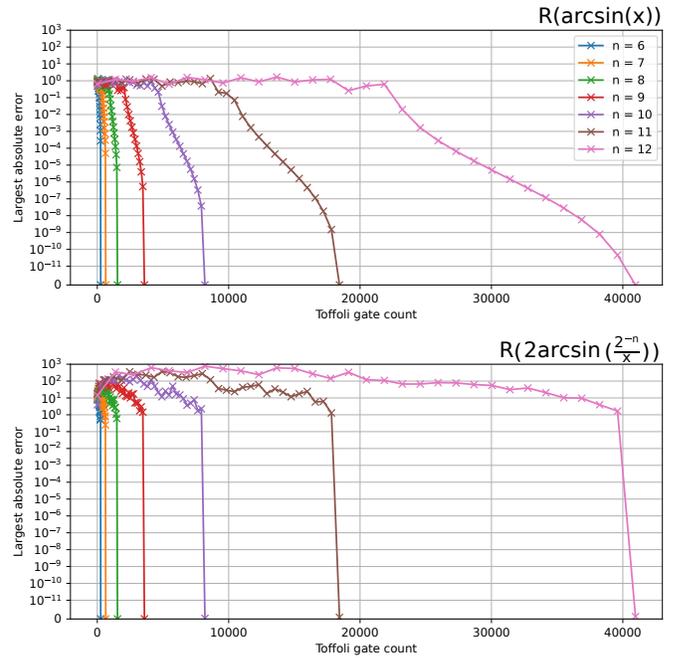}
    \caption{The accuracies of approximate lookup-tables compiled for the functions $\arcsin(x)$ (top), $2\arcsin\left( \frac{2^{-n}}{x}\right)$ (bottom) for different circuit depths (x-axis) and argument registers sizes $n$ (colored). Each circuit is evaluated on $2^n$ inputs in the interval $[-1, 1[$. Note that in the bottom plot the function $2\arcsin\left( \frac{2^{-n}}{x}\right)$ depends on $n$, which ensures that the function is defined for all inputs that are represented by the corresponding register.}
    \label{fig:approximation_efficiency_1_1}
\end{figure}

While the circuit approximation procedure is generally effective at reducing the circuit depths of non-polynomial functions, there are cases where it struggles to offer a satisfactory trade-off. The limitations of the approximation procedure are demonstrated in the simulations presented in \figurename~\ref{fig:approximation_efficiency_1_1}, where lookup-tables were compiled for the functions $R(\arcsin(x))$ and $R\left(2\arcsin\left( \frac{2^{-n}}{x} \right)\right)$, with circuits operating on values in the range $[-1, 1[$. Notably, the lookup-tables of \figurename~\ref{fig:approximation_efficiency_nonpols} evaluated for values in the same range exhibit steep slopes in the beginning of their graphs, while the lookup-tables of \figurename~\ref{fig:approximation_efficiency_1_1} do not. The arcsine function is notoriously difficult to approximate near the values of -1 and 1, and this is reflected in the highly irregular course of the graphs in \figurename~\ref{fig:approximation_efficiency_1_1}. For $R(\arcsin(x))$, it can be observed that the rotation can only be reasonably approximated up to a certain circuit depth, beyond which the circuit incurs large errors on the order of $10^{0}$. The case of $R\left(2\arcsin\left( \frac{2^{-n}}{x} \right)\right)$ is even more extreme. Unless all $2^n$ rotation gates are used, the circuit incurs extremely large errors, on the order of $10^3$. This effect arises because the rotation values of the gates in this circuit are large in absolute value relative to the range of the implemented function. This causes the circuit to alternate between over- and undershooting, rather than slowly approaching the correct result from one direction. As a result, the circuit is highly sensitive to omitting rotation gates. While this function would be especially useful in the context of the HHL algorithm, as it would absorb the eigenvalue inversion subroutine and reduce the overall circuit depth, it can not be approximated with our approach.

To summarize, while clearly limited to moderate register sizes and in terms of implementable functions, our approach allows compiling approximate circuits for many polynomial and non-polynomial functions with satisfactory accuracy, while needing only a few hundred Toffoli gates and almost no ancilla qubits.

\section{Conclusion}
\label{sec:conclusion}
The goal of this paper was to develop an efficient implementation of the arithmetic subroutines in the HHL algorithm, which represent a difficult task to perform on NISQ-devices. Known implementations to these problems typically require many gates and ancilla qubits. Classical computing methods often suffer from high ancilla requirements in the quantum realm as the reversible nature of quantum circuits requires intermediate results to be stored in ancillary registers. To address this issue, we investigated polynomial rotation circuits and lookup-tables, which are structurally simple and space-efficient but scale exponentially with the size of the argument register. We proposed a promising new method that trades off accuracy for reduced circuit depth, making these circuits feasible and competitive with other state-of-the-art methods. Our approach requires significantly fewer ancilla qubits than iterative methods such as ones that are based the on Horner scheme. The circuit depths can be reduced to only a few hundred Toffoli gates while providing good approximations with an accuracy of about $10^{-3}$ to $10^{-6}$. In addition, our approximation procedure maintains the structural simplicity of these circuits, which makes them very easy to specify and program in a circuit specification language of choice. 

While our approach is capable of providing approximate circuits for many polynomial and non-polynomial functions, it has limitations. On the one hand, there are functions for which the circuits cannot be approximated sufficiently well. For example, implementing circuits that evaluate the arcsine function over its entire domain still require a large number of Toffoli gates. In the worst case, certain functions can only be implemented with exponentially many gates, as an approximation would introduce very large errors. On the other hand, the exponential complexity of the compilation algorithms naturally limits the problem sizes for which these circuits can be compiled. Although classical computing provides significant computing power, compiling circuits for large register sizes would be time-consuming and impractical. It is also important to note that while we made a step towards facilitating NISQ-implementations for the HHL and similar algorithms, implementing circuits that require several hundred Toffoli gates will remain a challenge in the near future.

Despite these challenges, we believe that our approach has great potential for further improvement and optimization, which is facilitated by the structural and conceptual simplicity of the circuits. Future work on this topic could include exploring different implementations for multiple-controlled rotation gates, different heuristics for the contribution-to-cost ratio, or embedding the circuits in more sophisticated algorithms, e.g., by conditionally rotating around different functions in different intervals. Exploring these ideas could lead to further reduction in circuit size, an increase of accuracy or expanding the class of functions that can be approximated.

Overall, we believe that our proposed method offers a valuable contribution to the field of quantum computing, providing a promising alternative for implementing arbitrary function rotations with smaller circuit dimension, and paving the way for the implementation of resource-intensive algorithms on future quantum computers.



\section*{Acknowledgment}
This work was partially funded by the German BMWK project \textit{QCHALLenge} (01MQ22008A).

\bibliographystyle{unsrt}  
\bibliography{main} 

\end{document}